\documentclass[sigconf]{acmart}

\usepackage{fix-cm}
\copyrightyear{2026}
\acmYear{2026}
\setcopyright{cc}
\setcctype{by}
\acmConference[CCS '26]{Proceedings of the 2026 ACM SIGSAC Conference on Computer and Communications Security}{November 15--19, 2026}{The Hague, Netherlands}
\acmBooktitle{Proceedings of the 2026 ACM SIGSAC Conference on Computer and Communications Security (CCS '26), November 15--19, 2026, The Hague, Netherlands}
\acmDOI{10.1145/3830454.3846456}
\acmISBN{979-8-4007-2871-6/2026/11}

\usepackage{latexsym,fancyhdr,url}
\usepackage{enumitem}
\usepackage{amsmath}
\usepackage{booktabs}
\usepackage{nicefrac}
\usepackage{siunitx}
\usepackage{array,framed}
\usepackage{
  color,
  float,
  epsfig,
  wrapfig,
  graphics,
  graphicx,
  subcaption
}
\usepackage{textcomp}
\usepackage{setspace}
\usepackage{latexsym,fancyhdr,url}
\usepackage{algorithm2e}
\usepackage{algpseudocode}
\usepackage{xparse}
\usepackage{xspace}
\usepackage{multirow}
\usepackage{csvsimple}

\usepackage{
  tikz,
  pgfplots,
  pgfplotstable
}
\usepackage{hyperref}

\usetikzlibrary{
  shapes.geometric,
  arrows,
  external,
  pgfplots.groupplots,
  matrix
}

\pgfplotsset{compat=1.9}

\usepackage{mathtools}

\DeclareMathAlphabet{\mathcal}{OMS}{cmsy}{m}{n}

\newcommand{\SNR}{\mathrm{SNR}}
\newcommand{\SR}{\mathrm{SR}}
\newcommand{\GE}{\mathrm{GE}}
\newcommand{\Pinata}{\ensuremath{\mathrm{Pi\tilde{n}ata}}}

\DeclareGraphicsExtensions{%
    .png,.PNG,%
    .pdf,.PDF,%
    .jpg,.mps,.jpeg,.jbig2,.jb2,.JPG,.JPEG,.JBIG2,.JB2}

\usepackage{xparse}
\newcommand{\bnm}{\begin{newmath}}
\newcommand{\enm}{\end{newmath}}

\newcommand{\bea}{\begin{eqnarray*}}%
\newcommand{\eea}{\end{eqnarray*}}%

\newcommand{\bne}{\begin{newequation}}
\newcommand{\ene}{\end{newequation}}

\newcommand{\bal}{\begin{newalign}}
\newcommand{\eal}{\end{newalign}}

\newenvironment{newalign}{\begin{align}%
\setlength{\abovedisplayskip}{4pt}%
\setlength{\belowdisplayskip}{4pt}%
\setlength{\abovedisplayshortskip}{6pt}%
\setlength{\belowdisplayshortskip}{6pt} }{\end{align}}

\newenvironment{newmath}{\begin{displaymath}%
\setlength{\abovedisplayskip}{4pt}%
\setlength{\belowdisplayskip}{4pt}%
\setlength{\abovedisplayshortskip}{6pt}%
\setlength{\belowdisplayshortskip}{6pt} }{\end{displaymath}}

\newenvironment{newequation}{\begin{equation}%
\setlength{\abovedisplayskip}{4pt}%
\setlength{\belowdisplayskip}{4pt}%
\setlength{\abovedisplayshortskip}{6pt}%
\setlength{\belowdisplayshortskip}{6pt} }{\end{equation}}

\newcounter{ctr}

\newcounter{mytable}
\def\mytable{\begin{centering}\refstepcounter{mytable}}
\def\endmytable{\end{centering}}

\newcounter{myfig}
\def\myfig{\begin{centering}\refstepcounter{myfig}}
\def\endmyfig{\end{centering}}

\newlength{\saveparindent}
\newlength{\saveparskip}
\newcommand{\E}{{\rm I\kern-.3em E}}

\renewcommand{\eqref}[1]{\mbox{Equation~(\ref{#1})}}

\def \part {part}

\def \blackslug{\hbox{\hskip 1pt \vrule width 4pt height 8pt
    depth 1.5pt \hskip 1pt}}
\def \qed{\quad\blackslug\lower 8.5pt\null\par}

\newcounter{mynote}[section]

\newcommand\ignore[1]{}

\newcounter{rcnote}[section]

\newcounter{mrnote}[section]

\newcounter{fknote}[section]

\newcounter{anote}[section]

\DeclareMathSymbol{\mlq}{\mathord}{operators}{``}
\DeclareMathSymbol{\mrq}{\mathord}{operators}{`'}

\newcommand{\rhf}[2]{R_{f, \gamma}}

\DeclareDocumentCommand{\edist}{o o}{
  \ensuremath{
    \IfNoValueTF{#1}{{d}}{{\sf d}(#1,#2)}
  }
}

\newcommand{\olrk}[1]{\ifx\nursymbol#1\else\!\!\mskip4.5mu plus 0.5mu\left(\mskip0.5mu plus0.5mu #1\mskip1.5mu plus0.5mu \right)\fi}

\NewDocumentCommand{\indseq}{ O{1} O{r} }{{#1}\ldots {#2}}

\microtypesetup{nopatch=footnote}
\begin{document}

\fancyhead{}
\def\thetitle{Poster: s-MDM: Generative Virtualization of Multi-Device Hardware Variations for Portable DL-SCA}
\title{\thetitle}

\author{Niloufar Sayadi}
\orcid{0000-0003-4930-6556}
\correspondingauthor
\affiliation{%
  \institution{Centrum Wiskunde \& Informatica}
  \institution{Vrije Universiteit Amsterdam}
  \city{Amsterdam}
  \country{The Netherlands}
}
\email{n.sayadi@cwi.nl}

\author{Marten van Dijk}
\orcid{0000-0001-9388-8050}
\affiliation{%
  \institution{Centrum Wiskunde \& Informatica}
  \institution{Vrije Universiteit Amsterdam}
  \city{Amsterdam}
  \country{The Netherlands}
}
\email{marten.van.dijk@cwi.nl}

\author{Chenglu Jin}
\orcid{0000-0001-6306-8019}
\affiliation{%
  \institution{Centrum Wiskunde \& Informatica}
  \city{Amsterdam}
  \country{The Netherlands}
}
\email{chenglu.jin@cwi.nl}

\begin{abstract}
Deep Learning-based Side-Channel Analysis (DL-SCA) frequently suffers from catastrophic performance degradation across unseen hardware due to printed circuit board routing differences, silicon process variations, and measurement noise shifts. This poster presents the Synthetic Multiple Device Model (\mbox{s-MDM}), a zero-target-trace generative framework designed to improve cross-device portability. \mbox{s-MDM} combines a structured cVAE generator, a Walsh--Hadamard leakage anchor, continuous style modulation, and decoupled leakage--style--domain critics to synthesize virtual source-device profiles offline. Benchmarked on 32-bit side-channel traces (\texttt{AES\_PTv2}), \mbox{s-MDM} maps a precise operational boundary: while physical MDM remains superior on identical electrical clones ($D_4$), \mbox{s-MDM} achieves consistently low key rank on the layout/acquisition-shifted $\Pinata$ target, where physical baselines are unstable or misaligned. 

\end{abstract}

\keywords{Side-Channel Analysis, Deep Learning, Generative Modeling, Process Variation}

\ccsdesc[500]{Security and privacy~Hardware security implementation}
\ccsdesc[500]{Security and privacy~Side-channel analysis and countermeasures}
\ccsdesc[300]{Computing methodologies~Machine learning approaches}
\hypersetup{
  pdfauthor={Niloufar Sayadi, Marten van Dijk, Chenglu Jin},
  pdfkeywords={Side-Channel Analysis, Deep Learning, Generative Modeling, Process Variation}
}

\maketitle

\section{Introduction}
\label{sec:intro}
DL-SCA breaks cryptographic implementations efficiently~\cite{cagli2017convolutional}. However, classifiers suffer severe performance degradation across unseen target hardware due to the \textit{portability problem}: process variations, PCB routing differences, and noise shifts alter physical leakage distributions~\cite{bhasin2020mind, picek2023sok}. Existing approaches for cross-device portability have severe operational limits:
\begin{enumerate}[wide=0pt, noitemsep, topsep=0pt]
    \item \textbf{Physical Multi-Device Modeling (MDM):} MDM trains classifiers across a physical fleet of boards to learn fleet-invariant features and treat single-board traits as noise~\cite{bhasin2020mind}. However, procuring and profiling physical fleets incurs severe financial and logistical overheads~\cite{bhasin2020mind}.    
    \item \textbf{Target-Assisted Domain Adaptation:} Methods like CDPA~\cite{cao2021cross} and UDA fine-tuning~\cite{woo2025novel} align features across hardware, but strictly require capturing unlabeled target-device traces during profiling—a condition infeasible in non-permissive scenarios~\cite{karayalcin2024kind, picek2023sok}.
\end{enumerate}

\noindent\textbf{\emph{Our Contributions.}}
We present the \textbf{Synthetic Multiple Device Model (s-MDM)} for zero-target-trace \textit{device-efficient portability}. Our key contributions are:
\begin{itemize}[leftmargin=1.5em, noitemsep, topsep=0pt]
    \item We introduce a generative virtualization framework that constructs a continuous synthetic device manifold from minimal source hardware.
    \item We design a hybrid cVAE-GAN architecture that extends basis-projected modeling~\cite{zaid2023conditional, boussam2025optimal} with Gradient Reversal Layers (GRL), continuous style modulation, and decoupled critics to separate deterministic leakage from physical style variations.
    \item We empirically characterize the operational boundary between physical pooling and synthetic virtualization on \texttt{AES\_PTv2}, establishing when direct multi-device modeling remains preferable and when s-MDM improves generalization under layout and acquisition shift.
\end{itemize}

\noindent\textbf{\emph{Background and Related Work.}}
Portability methods following MDM~\cite{bhasin2020mind}, including Cross-Device Profiled Attacks (CDPA)~\cite{cao2021cross}, MMD minimization~\cite{woo2025novel}, and adversarial domain learning~\cite{cao2022al, yu2021cross}, align feature distributions between devices, but unlike MDM, they require unlabeled victim-device traces during profiling~\cite{cao2021cross, woo2025novel, picek2023sok}. Generative SCA methods have been used for augmentation~\cite{mukhtar2022fake}, explicit noise-leakage factorization over monomial bases (\mbox{cVAE-SA}~\cite{zaid2023conditional}, \mbox{cVAE-OSM}~\cite{boussam2025optimal}), or transferring white-box reference features to black-box targets (\mbox{CGAN-SCA}~\cite{karayalcin2024kind}). Existing generative SCA architectures mostly rely either on victim-device measurements for domain transfer~\cite{cao2022al, karayalcin2024kind} or restrict noise modeling to a single physical device without spanning distinct hardware configurations~\cite{zaid2023conditional, boussam2025optimal}. In contrast, s-MDM uses only source-device traces and synthesizes virtual device-style variation before deployment.

\section{Methodology}
\label{sec:methodology}
The s-MDM framework is engineered for zero-target-trace side-channel portability. Instead of relying on target-dependent domain adaptation, which requires unlabeled traces from the target device, s-MDM learns a generative model only from the available profiling devices and uses it offline to synthesize virtual device styles. The objective is to expose the downstream attack classifier to a wider range of physically plausible leakage conditions before deployment, while keeping the final target device completely unseen during training and validation.

\noindent\textbf{\emph{Leakage Decomposition.}}
s-MDM builds on cVAE-OSM \cite{boussam2025optimal} for structured leakage factorization. Given a raw trace observation vector $\mathbf{T} \in \mathbb{R}^{D}$, s-MDM models the trace through the following decomposition:
\begin{equation}
    \mathbf{T}
    =
    \Psi(Y)
    +
    \mathbf{N}_{\mathrm{content}}
    (\mathbf{z},\boldsymbol{\sigma}^{2})
    +
    \mathbf{N}_{\mathrm{style}}
    (\mathbf{z}_{\mathrm{style}})
    \label{eq:rigorous_smdm_decomp}
\end{equation}
\begin{itemize}[leftmargin=1.em, noitemsep, topsep=0pt]
    \sloppy
    \item $\Psi(Y) = b(Y)^{\!\top}\mathbf{W}_{\!\psi}$ is the deterministic AES-dependent leakage template, where $b(Y)$ is a 256-dimensional Walsh--Hadamard basis representation of the 8-bit intermediate value $Y$~\cite{boussam2025optimal}.
    \item $\mathbf{N}_{\mathrm{content}}(\mathbf{z}, \boldsymbol{\sigma}^{2})$ models stochastic content variation using a latent code $\mathbf{z}$ and heteroscedastic variance $\boldsymbol{\sigma}^{2}$.
    \item $\mathbf{N}_{\mathrm{style}}(\mathbf{z}_{\mathrm{style}})$ models a device-specific physical style signature through a continuous 16-dimensional style code.
\end{itemize}

\noindent\textbf{\emph{The Hybrid cVAE-GAN Architecture.}}
s-MDM implements the decomposition in
Eq.~\ref{eq:rigorous_smdm_decomp} through a hybrid cVAE--GAN. It couples the structured density modeling of a cVAE with the adversarial feedback of GAN-style critics to bypass traditional VAE waveform blurring. The generator disentangles execution content from style, while three decoupled critics ( $D_{\mathrm{style}}$, $D_{\mathrm{leak}}$, and $D_{\mathrm{domain}}$) enforce physical realism, cryptographic alignment, and device invariance. The proposed architecture comprises four core functional components:
\begin{enumerate}[wide=0pt, leftmargin=0.2cm,noitemsep, topsep=0pt]
    \item \textbf{Structured cVAE Generator:} Leveraging a cVAE-OSM backbone~\cite{boussam2025optimal}, the content encoder estimates $q_{\phi}(\mathbf{z}\mid \mathbf{T},b(Y))$ and heteroscedastic residual variance $\boldsymbol{\sigma}^{2}$. The deterministic subspace $\Psi(Y) = b(Y)^{\top}\mathbf{W}_{\psi}$ remains strictly anchored to prevent feature bleaching during optimization.
    \item \textbf{Continuous Style Modulation:} A blind, secret-unaware encoder extracts a style code $\mathbf{z}_{\mathrm{style}}$ capturing physical waveform appearance. The decoder re-renders the synthesized trace $\widehat{\mathbf{T}}$ by modulating the residual noise via Feature-wise Linear Modulation (FiLM). While Eq.~\ref{eq:rigorous_smdm_decomp} gives the conceptual trace decomposition, the implemented decoder realizes the style term through FiLM modulation of the residual content-noise pathway as
    \begin{equation}
        \widehat{\mathbf{T}} = \Psi(Y) + \boldsymbol{\gamma}(\mathbf{z}_{\mathrm{style}}) \odot \mathbf{N}_{\mathrm{content}}(\mathbf{z},\boldsymbol{\sigma}^{2}) + \boldsymbol{\beta}(\mathbf{z}_{\mathrm{style}}),
        \label{eq:film_decoder}
    \end{equation}
    where $\boldsymbol{\gamma}$ and $\boldsymbol{\beta}$ are trace-length scaling and shift vectors. Eq.~\ref{eq:film_decoder} explicitly realizes the style component from Eq.~\ref{eq:rigorous_smdm_decomp} as $N_{\mathrm{style}}(z_{\mathrm{style}}) = (\gamma(z_{\mathrm{style}})-1) \odot N_{\mathrm{content}} + \beta(z_{\mathrm{style}})$.
    \item \textbf{Decoupled Adversarial Critics:} $D_{\mathrm{style}}$ evaluates secret-blind physical trace morphology. Concurrently, the leakage critic $$D_{\mathrm{leak}}(\mathbf{T},Y) = \psi(\phi(\mathbf{T})) + \phi(\mathbf{T})^{\top}\mathbf{V}b(Y)$$ forces the CNN feature map $\phi(\mathbf{T})$ to preserve trace--label cryptographic alignment.
    \item \textbf{GRL Domain Safeguard:} A domain head $D_{\mathrm{domain}}$ predicts profiling board identity from $\phi(\mathbf{T})$. A Gradient Reversal Layer (GRL), acting as $\mathcal{R}_{\lambda}(\phi(\mathbf{T}))=\phi(\mathbf{T})$ with reversed backpropagation gradients ($\frac{\partial \mathcal{R}_{\lambda}}{\partial \phi} = -\lambda I$), penalizes board-identifying features and encourages device-invariant representations.
\end{enumerate}

\noindent\textbf{\emph{Continuous Manifold Virtualization.}}
Once the hybrid cVAE--GAN is trained on the available profiling fleet, the generator is frozen and used as an offline virtual-device synthesizer. Instead of sampling only from the observed profiling styles, s-MDM samples the continuous style code from an expanded latent envelope, $\mathbf{z}_{\mathrm{style}} \sim \mathcal{N}(0,\alpha\mathbf{I}), \quad \alpha = 1.5$, and combines $\mathbf{z}_{\mathrm{style}}$ with $Y$ and content latents via Eq.~\ref{eq:film_decoder}. This produces a virtual fleet of style-diverse traces that interpolate and mildly extrapolate beyond the observed source-board appearances without using any traces from the final target device. The purpose is to broaden the training distribution seen by the downstream attack classifier before deployment. The resulting dataset $\mathcal{D}_{\text{train}}^{\text{attack}} = \mathcal{D}_{\text{real}}^{\text{profiling}} \cup \mathcal{D}_{\text{synthetic}}^{\text{s-MDM}}$ trains a downstream attack classifier $C_{\theta}$ offline. During the final attack (key recovery), target traces are processed directly by the trained classifier while the generator and all adversarial critics are discarded:
\[
    \mathbf{T}_{\mathrm{target}}
    \longrightarrow
    C_{\theta}(\mathbf{T}_{\mathrm{target}})
    \longrightarrow
    p(Y\mid \mathbf{T}_{\mathrm{target}})
    \longrightarrow
    \text{key ranking}.
\]    
No target traces are used to train, adapt, validate, or augment the generator. To avoid contaminating the sealed target evaluation, s-MDM uses a Leave-One-Board-Out (LOBO) validation protocol. The model is trained on the source boards and validated on a held-out board from the source-device family that acts as a pseudo-target. This held-out board provides a portability proxy for selecting training checkpoints and monitoring leakage-fidelity diagnostics such as $\SNR_\Psi$, guessing entropy, and success rate. The final target remains completely unseen until the final deployment evaluation.
\section{Experimental Evaluations}
\label{sec:eval}
\noindent\textbf{\emph{Experimental Setup.}}
The s-MDM framework was evaluated on an unprotected 32-bit software AES-128 engine using the \href{https://github.com/urioja/AESPTv2}{\texttt{AES\_PTv2}} side-channel trace infrastructure. 
The device fleet contains four nominally identical STM32F411E-DISCO boards ($D_1$--$D_4$) and one distinct Riscure $\Pinata$ board. Boards $D_1$--$D_4$ use the STM32F411VE Cortex-M4 platform and are measured through a noisy, less-invasive EM setup, while $\Pinata$ is a Cortex-M4F-based training board with a modified power network and cleaner direct power acquisition. We use $D_1$ and $D_2$ for profiling, $D_3$ for LOBO validation, $D_4$ as an unseen electrical clone target, and $\Pinata$ as a layout/acquisition-shifted target.

Each device is characterized by its leakage-peak sample index, peak \(\SNR\), \(\rho=\SNR/(1+\SNR)\), and device-specific characterization window in Table~\ref{tab:devices}. For model ingestion, all traces are cropped around each device's leakage peak to a common 120-sample width. We report guessing entropy \(\GE\), the mean zero-based rank of the correct key, and success rate \(\SR\), the fraction of runs reaching rank zero. While the profiling boards are strongly noise-dominated ($\rho_{D_1}=0.072$), the \Pinata{} target shifts from the noise-dominated profiling regime to a signal-dominated regime (\(\rho_{\Pinata}=0.737\)), consistent with a substantial layout/acquisition shift rather than ordinary clone-to-clone variation.
\begin{table}[htbp]
\centering
\footnotesize
\caption{Physical characterization of the device fleet.}
\label{tab:devices}
\renewcommand{\arraystretch}{0.80} 
\setlength{\tabcolsep}{5pt} 
\resizebox{\linewidth}{!}{%
\begin{tabular}{@{}l c c c c l@{}}
\toprule
\textbf{Device} & \textbf{Role} & \textbf{Peak idx.} & \textbf{Peak $\SNR$} & \textbf{$\rho$} & \textbf{Device-optimal window} \\
\midrule
$D_1$    & profiling        & 48  & 0.077 & 0.072 & $[0,128)$   \\
$D_2$    & profiling        & 63  & 0.135 & 0.119 & $[0,143)$   \\
$D_3$    & validation       & 39  & 0.260 & 0.206 & $[0,119)$   \\
$D_4$    & sealed target    & 39  & 0.224 & 0.183 & $[0,119)$   \\
$\Pinata$   & sealed target    & 104 & \textbf{2.796} & \textbf{0.737} & $[24,184)$ \\
\bottomrule
\end{tabular}}
\end{table}

\noindent\textbf{\emph{In-Manifold Evaluation (Electrical Clone).}}
On the electrical clone target $D_4$ ($\SNR=0.224$, $\rho_{D_4}=0.183$), physical pooling is optimal: Physical MDM reaches full recovery in 109 traces and the two-device baseline in 586 traces (Table~\ref{tab:d4_results}). In contrast, our s-MDM improves over single-device training but does not converge within the evaluated trace budget, as shown in Table~\ref{tab:d4_results}. 
This result is consistent with the clone setting where real multi-device traces are more informative than synthetic re-rendering. In this regime, the generator can act as a compression bottleneck rather than a portability advantage.
\begin{table}[htbp]
\centering
\footnotesize
\caption{Portability Profiles on Electrical Clone Target $D_4$.}
\label{tab:d4_results}
\renewcommand{\arraystretch}{0.80}
\resizebox{\linewidth}{!}{%
\begin{tabular}{@{}l c c c c@{}}
\toprule
\textbf{Strategy} & \textbf{Physical Support} & \textbf{Final $\GE$} & \textbf{Final $\SR$} & \textbf{Conv. (Traces)} \\
\midrule
Single-Device    & 1 ($D_1$) & 18.36 & 0.16 & --- \\
Two-Device       & 2 ($D_1, D_2$) & 0.00 & 1.00 & 586 \\
Physical MDM     & 3 ($D_1$--$D_3$) & 0.00 & 1.00 & 109 \\
s-MDM (Full)     & Virtualized (from 2) & 5.72 & 0.21 & --- \\
\bottomrule
\end{tabular}%
}
\end{table}    

\noindent\textbf{\emph{Out-of-Manifold Evaluation (Layout and Acquisition Shift: \Pinata).}}
The benefit of s-MDM appears in the out-of-support setting. $\Pinata$ has a much stronger leakage regime than profiling boards (\(\SNR=2.796\), \(\rho_{\Pinata}=0.737\)).
\balance

Table~\ref{tab:pinata_results} shows that the physical baselines are unstable under this shift. Single-device training fails to recover, two-device training is highly seed-sensitive, and Physical MDM ranks the key worse than random. In contrast, s-MDM achieves low guessing entropy, with $\mathrm{GE}\leq3.3$ across the three independent realizations.
The improvement suggests that continuous style virtualization exposes the classifier to a broader range of leakage appearances before deployment, reducing its dependence on the low-SNR morphology of the profiling boards.
\begin{table}[htbp]
\centering
\footnotesize
\caption{Replication of final $\mathrm{GE}$ across three independent random seeds on the $\Pinata$ target ($\mathrm{SNR}=2.796$, $\rho_{\Pinata}=0.737$). $\mathrm{GE}_{\text{rand}}=127.5$ (256 keys); higher values indicate anti-correlated (worse-than-random) ranking.}
\label{tab:pinata_results}
\footnotesize
\begin{tabular*}{\linewidth}{@{\extracolsep{\fill}}l c c c c@{}}
\toprule
\textbf{Strategy} & \textbf{Seed 0} & \textbf{Seed 1} & \textbf{Seed 2} & \textbf{Status} \\
\midrule
Single-Device        & 36.32  & 18.00  & 16.95  & Fails \\
Two-Device           & 25.50  & 0.98   & 235.47 & Unstable \\
Physical MDM         & 146.36 & 239.07 & 243.79 & Anti-Correlated \\
\textbf{s-MDM (Full)} & \textbf{0.42} & \textbf{0.69} & \textbf{3.28} & \textbf{Recovers} \\
\bottomrule
\end{tabular*}
\end{table}

\noindent\textbf{\emph{Ablation Study.}}
To identify which architectural component drives the observed transfer, we ablate the main components of s-MDM on the layout-shifted target $\Pinata$, as detailed in Table~\ref{tab:ablation_results}. Each ablation reflects a single training run. For calibration, the full configuration spans $\mathrm{GE}=0.42$--$3.28$ across the three independent random seeds in Table~\ref{tab:pinata_results}. Only removal of continuous style diversity ($\mathrm{GE}=7.39$) degrades performance beyond this range, reducing the \(\SR\) from $0.71$ to $0.33$. Thus, style diversity is the only component whose removal shows a resolved performance degradation under the observed run-to-run variability.
\begin{table}[htbp]
\centering
\footnotesize
\caption{Structural ablation of generative components ($\Pinata$ target, single run per configuration). For reference, the full configuration spans $\mathrm{GE}=0.42$--$3.28$ across the seeds of Table~\ref{tab:pinata_results}.}
\label{tab:ablation_results}
\renewcommand{\arraystretch}{0.80} 
\resizebox{\linewidth}{!}{%
\begin{tabular}{@{}l c c c@{}}
\toprule
\textbf{Ablation State} & \textbf{Final $\GE$} & \textbf{Final $\SR$} & \textbf{Fidelity ($\SNR_{\Psi}$)} \\
\midrule
s-MDM Complete               & 0.42 & 0.71 & 0.057 \\
Without Style Diversity     & \textbf{7.39} & \textbf{0.33} & 0.058 \\
Without Device-Invariance   & 0.56 & 0.81 & 0.058 \\
Without Adversarial Critics & 0.03 & 0.97 & 0.055 \\
Without Subspace Anchor     & 0.43 & 0.80 & 0.057 \\
\bottomrule
\end{tabular}%
}
\end{table}
The remaining ablations fall inside the observed seed spread and are therefore not individually resolved here. In particular, although removing the adversarial critics yields $\mathrm{GE}=0.03$ in this run, this result cannot establish a definitive improvement over the complete model without multi-seed replication.

\section{Conclusion and Future Work}
\label{sec:conclusion}
\mbox{s-MDM} demonstrates that zero-target-trace generative virtualization can improve DL-SCA portability bottlenecks under severe layout and acquisition shifts. On the electrical clone target $D_4$, physical MDM remains superior, showing that synthetic re-rendering acts as an information filter when the target lies inside the source manifold. On \textsf{Pi\~{n}ata}, \mbox{s-MDM} reaches low \(\GE\) while physical baselines are unstable or anti-correlated. The ablation study identifies continuous style diversity as the only component whose removal causes a resolved performance degradation; the roles of the critics require multi-seed replication. Future work will refine critic optimization, calibrate style priors, and evaluate larger multi-board fleets.

\section*{Acknowledgments}
Marten van Dijk and Chenglu Jin are (partially) supported by project CiCS of the research programme Gravitation, which is (partly) financed by the Dutch Research Council (NWO) under the grant 024.006.037.


\begin{thebibliography}{11}


\ifx \showCODEN    \undefined \def \showCODEN     #1{\unskip}     \fi
\ifx \showISBNx    \undefined \def \showISBNx     #1{\unskip}     \fi
\ifx \showISBNxiii \undefined \def \showISBNxiii  #1{\unskip}     \fi
\ifx \showISSN     \undefined \def \showISSN      #1{\unskip}     \fi
\ifx \showLCCN     \undefined \def \showLCCN      #1{\unskip}     \fi
\ifx \shownote     \undefined \def \shownote      #1{#1}          \fi
\ifx \showarticletitle \undefined \def \showarticletitle #1{#1}   \fi
\ifx \showURL      \undefined \def \showURL       {\relax}        \fi
\providecommand\bibfield[2]{#2}
\providecommand\bibinfo[2]{#2}
\providecommand\natexlab[1]{#1}
\providecommand\showeprint[2][]{arXiv:#2}

\bibitem[Bhasin et~al\mbox{.}(2020)]%
        {bhasin2020mind}
\bibfield{author}{\bibinfo{person}{Shivam Bhasin}, \bibinfo{person}{Anupam Chattopadhyay}, \bibinfo{person}{Annelie Heuser}, \bibinfo{person}{Dirmanto Jap}, \bibinfo{person}{Stjepan Picek}, {and} \bibinfo{person}{Ritu Ranjan~Shrivastwa}.} \bibinfo{year}{2020}\natexlab{}.
\newblock \showarticletitle{Mind the portability: A warriors guide through realistic profiled side-channel analysis}. In \bibinfo{booktitle}{\emph{NDSS 2020-Network and Distributed System Security Symposium}}. \bibinfo{publisher}{Internet Society}, \bibinfo{address}{Reston, VA}, \bibinfo{pages}{1--14}.
\newblock


\bibitem[Boussam et~al\mbox{.}(2025)]%
        {boussam2025optimal}
\bibfield{author}{\bibinfo{person}{Sana Boussam}, \bibinfo{person}{Mathieu Carbone}, \bibinfo{person}{Beno{\^\i}t G{\'e}rard}, \bibinfo{person}{Gu{\'e}na{\"e}l Renault}, {and} \bibinfo{person}{Gabriel Zaid}.} \bibinfo{year}{2025}\natexlab{}.
\newblock \showarticletitle{Optimal Dimensionality Reduction using Conditional Variational AutoEncoder}.
\newblock \bibinfo{journal}{\emph{IACR Transactions on Cryptographic Hardware and Embedded Systems}} \bibinfo{volume}{2025}, \bibinfo{number}{3} (\bibinfo{year}{2025}), \bibinfo{pages}{164--211}.
\newblock


\bibitem[Cagli et~al\mbox{.}(2017)]%
        {cagli2017convolutional}
\bibfield{author}{\bibinfo{person}{Eleonora Cagli}, \bibinfo{person}{C{\'e}cile Dumas}, {and} \bibinfo{person}{Emmanuel Prouff}.} \bibinfo{year}{2017}\natexlab{}.
\newblock \showarticletitle{Convolutional Neural Networks with Data Augmentation Against Jitter-Based Countermeasures}. In \bibinfo{booktitle}{\emph{Cryptographic Hardware and Embedded Systems (CHES)}}. \bibinfo{publisher}{Springer}, \bibinfo{address}{Taipei, Taiwan}, \bibinfo{pages}{45--68}.
\newblock


\bibitem[Cao et~al\mbox{.}(2021)]%
        {cao2021cross}
\bibfield{author}{\bibinfo{person}{Pei Cao}, \bibinfo{person}{Chi Zhang}, \bibinfo{person}{Xiangjun Lu}, {and} \bibinfo{person}{Dawu Gu}.} \bibinfo{year}{2021}\natexlab{}.
\newblock \showarticletitle{Cross-Device Profiled Side-Channel Attack with Unsupervised Domain Adaptation}.
\newblock \bibinfo{journal}{\emph{IACR Transactions on Cryptographic Hardware and Embedded Systems (TCHES)}} \bibinfo{volume}{2021}, \bibinfo{number}{4} (\bibinfo{year}{2021}), \bibinfo{pages}{27--56}.
\newblock


\bibitem[Cao et~al\mbox{.}(2022)]%
        {cao2022al}
\bibfield{author}{\bibinfo{person}{Pei Cao}, \bibinfo{person}{Hongyi Zhang}, \bibinfo{person}{Dawu Gu}, \bibinfo{person}{Yan Lu}, {and} \bibinfo{person}{Yidong Yuan}.} \bibinfo{year}{2022}\natexlab{}.
\newblock \showarticletitle{AL-PA: Cross-Device Profiled Side-Channel Attack Using Adversarial Learning}. In \bibinfo{booktitle}{\emph{Proceedings of the 59th ACM/IEEE Design Automation Conference (DAC)}}. \bibinfo{publisher}{ACM}, \bibinfo{address}{San Francisco, CA, USA}, \bibinfo{pages}{691--696}.
\newblock


\bibitem[Karayalçın et~al\mbox{.}(2024)]%
        {karayalcin2024kind}
\bibfield{author}{\bibinfo{person}{Sengim Karayalçın}, \bibinfo{person}{Marina Kr{\v{c}}ek}, \bibinfo{person}{Lichao Wu}, \bibinfo{person}{Stjepan Picek}, {and} \bibinfo{person}{Guilherme Perin}.} \bibinfo{year}{2024}\natexlab{}.
\newblock \showarticletitle{It's a Kind of Magic: A Novel Conditional GAN Framework for Efficient Profiling Side-Channel Analysis}. In \bibinfo{booktitle}{\emph{Advances in Cryptology -- ASIACRYPT 2024}}. \bibinfo{publisher}{Springer}, \bibinfo{address}{Kolkata, India}, \bibinfo{pages}{99--131}.
\newblock


\bibitem[Mukhtar et~al\mbox{.}(2022)]%
        {mukhtar2022fake}
\bibfield{author}{\bibinfo{person}{Naila Mukhtar}, \bibinfo{person}{Lejla Batina}, \bibinfo{person}{Stjepan Picek}, {and} \bibinfo{person}{Yinan Kong}.} \bibinfo{year}{2022}\natexlab{}.
\newblock \showarticletitle{Fake It Till You Make It: Data Augmentation Using Generative Adversarial Networks for All the Crypto You Need on Small Devices}. In \bibinfo{booktitle}{\emph{CT-RSA 2022}}. \bibinfo{publisher}{Springer}, \bibinfo{address}{San Francisco, CA}, \bibinfo{pages}{297--321}.
\newblock


\bibitem[Picek et~al\mbox{.}(2023)]%
        {picek2023sok}
\bibfield{author}{\bibinfo{person}{Stjepan Picek}, \bibinfo{person}{Guilherme Perin}, \bibinfo{person}{Luca Mariot}, \bibinfo{person}{Lichao Wu}, {and} \bibinfo{person}{Lejla Batina}.} \bibinfo{year}{2023}\natexlab{}.
\newblock \showarticletitle{SoK: Deep Learning-Based Physical Side-Channel Analysis}.
\newblock \bibinfo{journal}{\emph{Comput. Surveys}} \bibinfo{volume}{55}, \bibinfo{number}{11} (\bibinfo{year}{2023}), \bibinfo{pages}{1--35}.
\newblock


\bibitem[Woo et~al\mbox{.}(2025)]%
        {woo2025novel}
\bibfield{author}{\bibinfo{person}{Ji-Eun Woo}, \bibinfo{person}{Yongsung Jeon}, \bibinfo{person}{Ju-Hwan Kim}, {and} \bibinfo{person}{Dong-Guk Han}.} \bibinfo{year}{2025}\natexlab{}.
\newblock \showarticletitle{Novel deep learning-based side-channel attack on different-device}.
\newblock \bibinfo{journal}{\emph{Soft Computing}} \bibinfo{volume}{29}, \bibinfo{number}{8} (\bibinfo{year}{2025}), \bibinfo{pages}{3847--3854}.
\newblock


\bibitem[Yu et~al\mbox{.}(2021)]%
        {yu2021cross}
\bibfield{author}{\bibinfo{person}{Honggang Yu}, \bibinfo{person}{Haoqi Shan}, \bibinfo{person}{Maximillian Panoff}, {and} \bibinfo{person}{Yier Jin}.} \bibinfo{year}{2021}\natexlab{}.
\newblock \showarticletitle{Cross-Device Profiled Side-Channel Attacks Using Meta-Transfer Learning}. In \bibinfo{booktitle}{\emph{ACM/IEEE Design Automation Conference (DAC)}}. \bibinfo{publisher}{IEEE}, \bibinfo{address}{San Francisco, CA}, \bibinfo{pages}{703--708}.
\newblock


\bibitem[Zaid et~al\mbox{.}(2023)]%
        {zaid2023conditional}
\bibfield{author}{\bibinfo{person}{Gabriel Zaid}, \bibinfo{person}{Lilian Bossuet}, \bibinfo{person}{Mathieu Carbone}, \bibinfo{person}{Amaury Habrard}, {and} \bibinfo{person}{Alexandre Venelli}.} \bibinfo{year}{2023}\natexlab{}.
\newblock \showarticletitle{Conditional variational autoencoder based on stochastic attacks}.
\newblock \bibinfo{journal}{\emph{IACR TCHES}} \bibinfo{volume}{2023}, \bibinfo{number}{2} (\bibinfo{year}{2023}), \bibinfo{pages}{310--357}.
\newblock


\end{thebibliography}

\end{document}